\documentclass[preprint2]{aastex}

\shorttitle{Prominence visibility in SXR}
\shortauthors{Schwartz et al.}

\usepackage{graphicx}
\usepackage{natbib}
\usepackage{url}
\usepackage{color}

\begin{document}


\title{PROMINENCE VISIBILITY IN Hinode/XRT IMAGES}



\author{P. Schwartz}
\affil{Astronomical Institute of Slovak Academy of Sciences, \\ 05960 Tatransk\'{a} Lomnica, Slovak Republic}
\email{pschwartz@astro.sk}

\author{S. Jej\v{c}i\v{c}}
\affil{University of Ljubljana, Faculty of Mathematics and Physics, 1000 Ljubljana, Slovenia}

\author{P. Heinzel}
\affil{Astronomical Institute, Czech Academy of Sciences, \\ 25165 Ond\v{r}ejov, Czech Republic}

\author{U. Anzer}
\affil{Max-Planck-Institut f\"{u}r Astrophysik, Karl-Schwarzschild-Strasse 1, \\ 85740 Garching, Germany}

\author{P.R. Jibben}
\affil{Harvard-Smithsonian Center for Astrophysics, 60 Garden Street,\\ 
             Cambridge, MA 02138, USA}




\begin{abstract}
In this paper we study the soft X\discretionary{-}{-}{-}ray (SXR) signatures of one particular
prominence. The X\discretionary{-}{-}{-}ray observations used here were made by the Hinode/XRT 
instrument using two different filters. Both of them have a
pronounced peak of the response function around 10 \AA. One of
them has a secondary smaller peak around 170 \AA, which leads to
a contamination of SXR images. The observed darkening in both
of these filters has a very large vertical extension. The position and
shape of the darkening corresponds nicely with the prominence
structure seen in SDO/AIA images. First we have investigated the
possibility that the darkening is caused by X-ray absorption. But
detailed calculations of the optical thickness in this spectral range 
show clearly that this effect is completely negligible. Therefore the
alternative is the presence of an extended region with a large 
emissivity deficit which can be caused by the presence of cool prominence plasmas
within otherwise hot corona. 
To reproduce the observed darkening one needs a very large extension 
along the line-of-sight of the region amounting to around 10$^5$ km. 
We interpret this region as the prominence spine, which is also 
consistent with SDO/AIA observations in EUV.

\end{abstract}




\newlength{\origacs}
\setlength\origacs{\the\arraycolsep}
\section{Introduction}
\label{s:intro}
Solar prominences observed above the limb are typically seen in emission against the
dark coronal background. This is the case of monochromatic imaging in spectral lines
formed at low temperatures like e.g. the hydrogen H$\alpha$ line or transition\discretionary{-}{-}{-}region 
spectral lines formed at temperatures of the prominence\discretionary{-}{-}{-}corona transition region (PCTR). 
In the latter case we see bright UV or EUV
prominences still against the dark corona which is not emitting in such lines. However, at
coronal temperatures, highly ionised atoms  emit radiation in various lines of different
species and we thus see the bright corona extending to large altitudes. In lines which have
 wavelengths below the Lyman limit of the neutral hydrogen (912\,\AA), we can often see
prominences as dark structures against such bright coronal background. This 'reversed'
visibility of prominences in EUV coronal lines is mainly caused by the absorption of the background coronal
radiation by cool hydrogen and helium plasma, where the neutral hydrogen (\ion{H}{1}), neutral
helium (\ion{He}{1}) and singly ionised helium (\ion{He}{2}) are photoionised at wavelengths below 912\,\AA, 
504\,\AA\ and 228\,\AA, respectively, depending on the wavelength of the coronal line under
consideration. For the limb prominences, this was quantitatively studied by \citet{cit:kucera98}
and later by several other authors. The photoionisation process is detailed in \citet{cit:ah2005}
who also described an additional mechanism of EUV prominence darkening.
The later was initially called emissivity blocking, but in
\citet{schwartz2014} the more 
appropriate term {\em emissivity deficit} is introduced 
since the blocking may evoke the situation when the background coronal radiation is somehow obscured 
by the prominence which is actually the case of the photoionisation absorption described above. 
Therefore we will continue in using of the term `emissivity depression' also in this work. 

Many nice examples of dark EUV prominence structures, both quiescent as well as eruptive, 
have been detected by by the \textit{Atmospheric Imaging Assembly} (AIA , \citeauthor{cit:aia}\ 
\citeyear{cit:aia}) EUV imager on board of the \textit{Solar Dynamics observatory} (SDO) satellite, or
\textit{Extreme Ultra-Violet Imager} (EUVI, \citeauthor{cit:euvi}\ \citeyear{cit:euvi}) instrument 
of the SECCHI instrument suite on\discretionary{-}{-}{-}board 
of the \textit{Solar Terrestrial Relations Observatory} (STEREO, 
\citeauthor{cit:stereo}\ \citeyear{cit:stereo}) satellites.  
Similar observations were made in earlier times also by the 
\textit{Extreme ultraviolet Imaging Telescope} (EIT,\ \citeauthor{cit:eit}\ \citeyear{cit:eit}) onboard the 
\textit{Solar and Heliospheric Observatory} (SOHO) satellite 
or \textit{Transition Region and Coronal Explorer} (TRACE, see \url{http://trace.lmsal.com}).  

Prominences are also seen in rasters of the \textit{EUV Imaging Spectrometer} 
(EIS, \citeauthor{cit:eis}\ \citeyear{cit:eis}) onboard the Hinode satellite \citep{cit:hinode}. 
The natural question then arises how far in EUV 
wavelengths can we detect such absorption and/or emissivity depression. This was studied by \citet{cit:anzeretal2007}  
who used the SOHO/EIT images of a quiescent prominence, together with soft X\discretionary{-}{-}{-}ray images obtained
by \textit{Soft X\discretionary{-}{-}{-}ray Telescope} (SXT, \citeauthor {cit:sxt}\ \citeyear{cit:sxt}) 
onboard the Yohkoh satellite. While in the 171\,\AA\ and 193\,\AA\  EIT channels 
the prominence was clearly visible as a dark absorbing structure, the co\discretionary{-}{-}{-}aligned SXT
image shows no signature of such darkening, perhaps except a weak visibility of the coronal
cavity surrounding the prominence. Therefore, these authors have concluded that there is a
negligible absorption at wavelengths around 50\,\AA\ where the SXT image was taken. This
was confirmed by numerical estimates performed according to \citet{cit:ah2005},
under typical prominence conditions. Nevertheless, the emissivity depression can not be
excluded and in fact it was used by \citet{heinzel2008apj} and by \citet{schwartz2014} for
analysis of dark features on the limb where both the prominence and cavity were observed.

Going to shorter X\discretionary{-}{-}{-}ray wavelengths around 10\,\AA, the Hinode/XRT has surprisingly revealed
dark prominence features quite similar to those visible in EUV. It is therefore the aim of the present
study to understand the nature of those SXR structures.  
We first consider the absorption of coronal SXR radiation by cool prominence plasmas, although
this was shown to be negligible around 50\,\AA\ where only hydrogen and helium was considered
\citep{cit:anzeretal2007}. However, at the XRT X\discretionary{-}{-}{-}ray wavelength range where the 
transmittance peaks around 10\,\AA, the absorption is much more complex. This was considered by various 
authors who demonstrated the importance of soft X\discretionary{-}{-}{-}ray absorption for heating of the 
solar chromosphere and chromospheric flare ribbons 
\citep{henouxandnakagawa1977, cit:hawleyandfisher1992, cit:berlickiandheinzel2004}. 
The absorption below 50\,\AA\, is enhanced, or even dominated, by various metals - for stellar applications
see e.g. \citet{cit:lon81}. A presence of such kind of absorption in prominences, if any, could thus
play a role in their energetics. We thus carefully compute the absorption by hydrogen, helium and important 
metals under typical prominence conditions in this study. We also provide a first observational evidence 
of the emissivity deficit effect. 
The paper is organised as follows: The SXR and EUV observations of a quiescent 
prominence are described in the next section and in section \ref{s:xrayvisibility} its visibility 
in SXR images taken by XRT is shown. In section \ref{s:mechsxrdark} and its subsections, 
three different mechanisms possibly leading to visibility of the prominence in XRT
images are studied and results are compared with observations. 
Section \ref{s:discussion} gives the discussion and our conclusions. 
\section{Observations}
\label{s:observations}
\begin{figure*}
\centering
\parbox{0.48\hsize}{
\resizebox{\hsize}{!}{\includegraphics{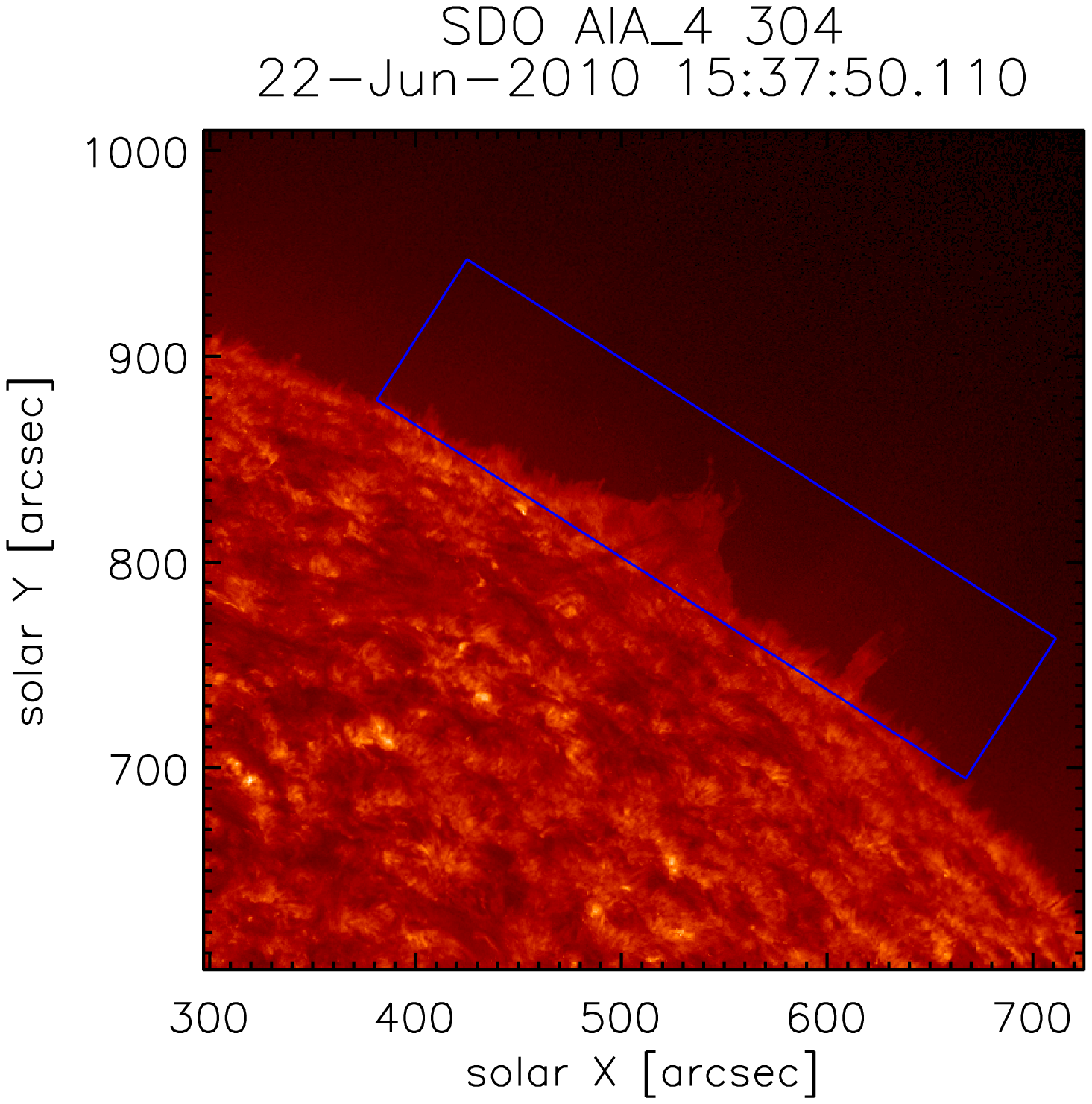}}}\
\parbox{0.48\hsize}{
\resizebox{\hsize}{!}{\includegraphics{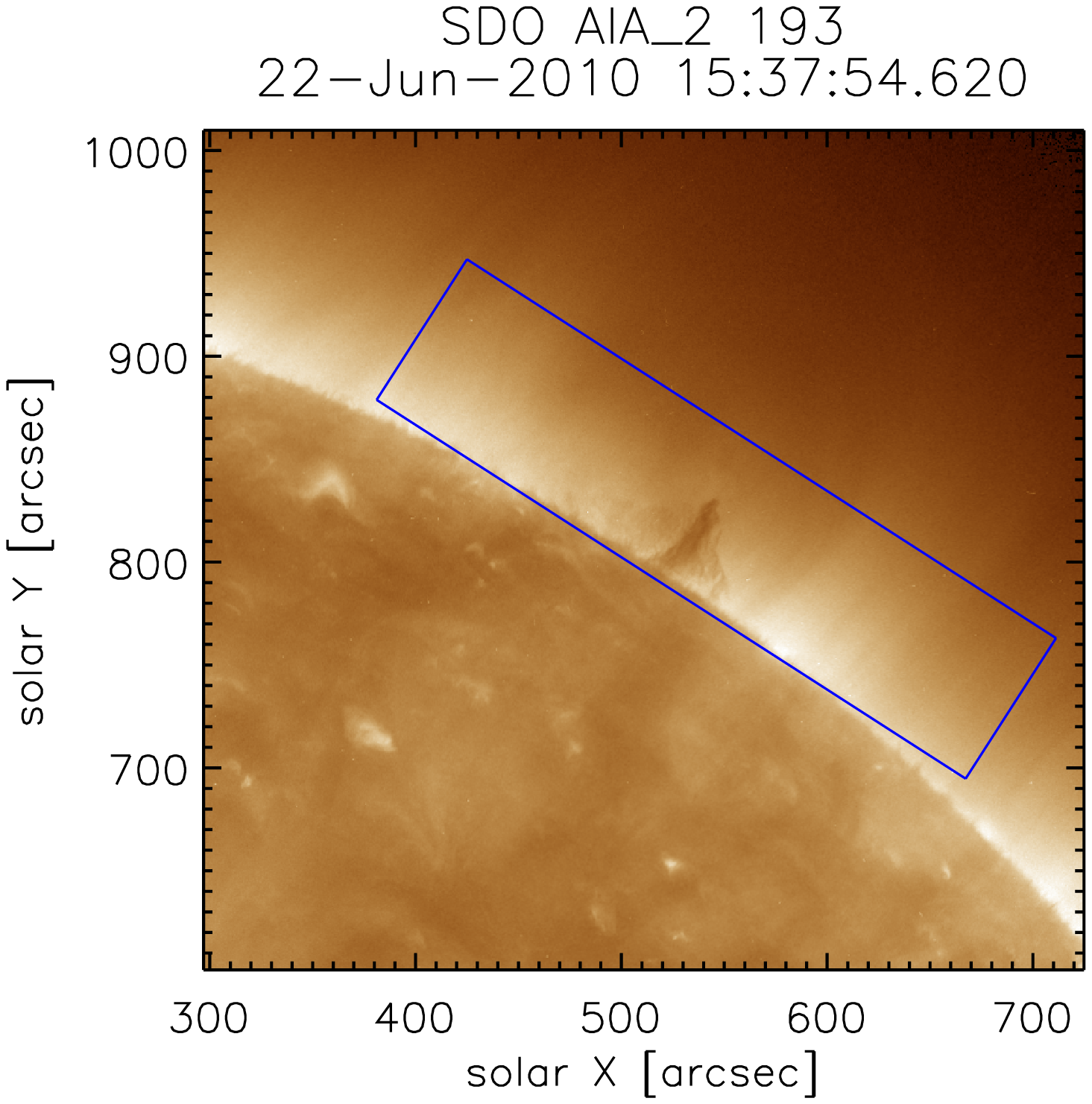}}}\\
\parbox{0.48\hsize}{
\resizebox{\hsize}{!}{\includegraphics{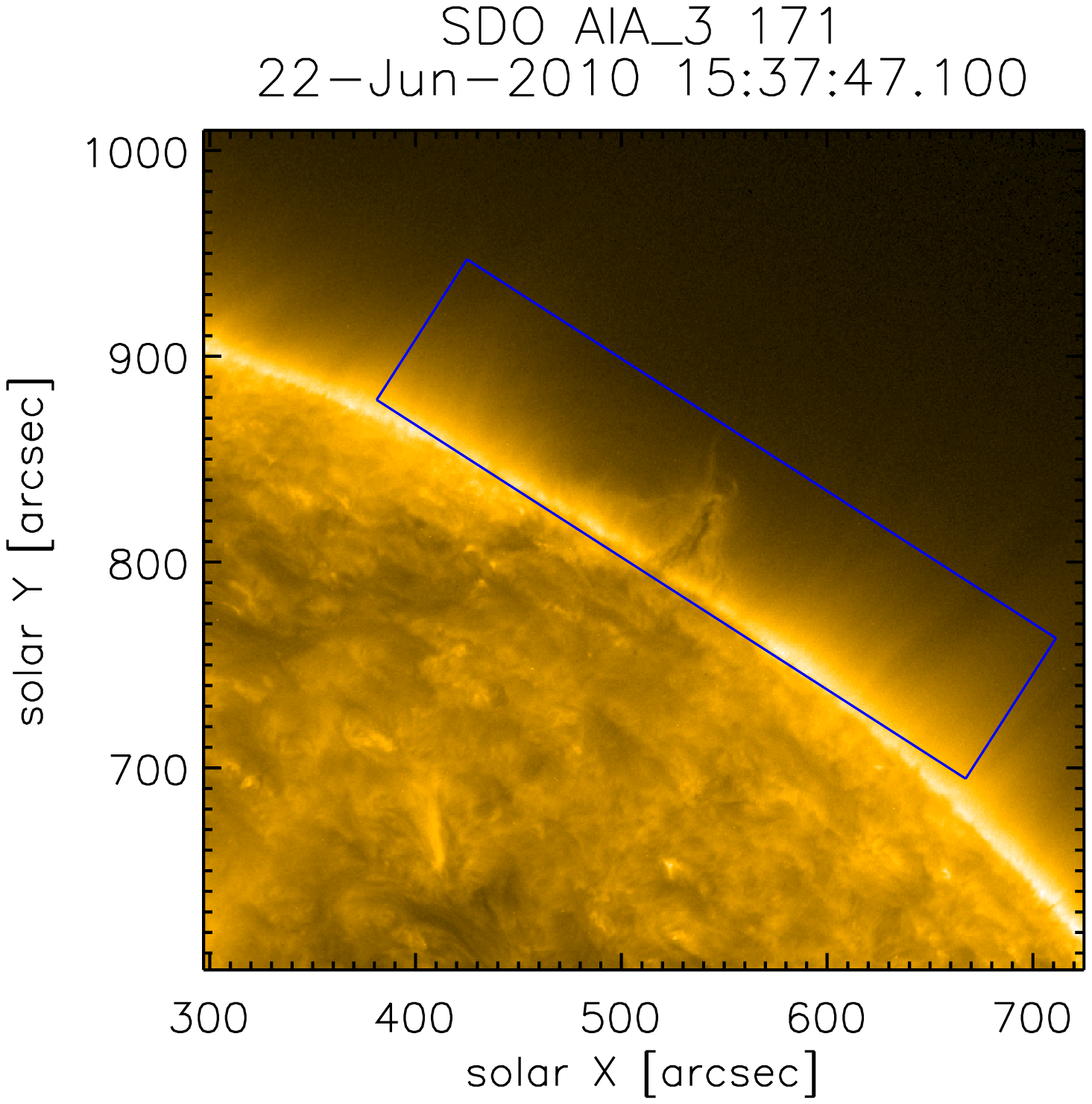}}}
\caption{Prominence observations made by the AIA instrument are shown in 
its three wavelength channels: 304\,\AA\ (the left upper panel) dominated by 
the \ion{He}{2} Lyman $\alpha$ line, 193\,\AA\ (the right upper panel) 
where mainly radiation of the \ion{Fe}{12} and \ion{Fe}{24} lines are  
detected and 171\,\AA\ (the lower panel) where the \ion{Fe}{9} and 
\ion{Fe}{10} lines contribute. In the 304\,\AA\ channel image whole prominence is seen well 
in emission while in 193\,\AA\ and 171\,\AA\ images mainly its spine is seen as dark structure. 
Prominence barbs are seen in emission in the 171\,\AA\ channel due to the \ion{Fe}{9} line 
formed in PCTR.} 
\label{fig:aiaobs1}
\end{figure*}

\begin{figure}
\begin{center}
\resizebox{\hsize}{!}{\includegraphics{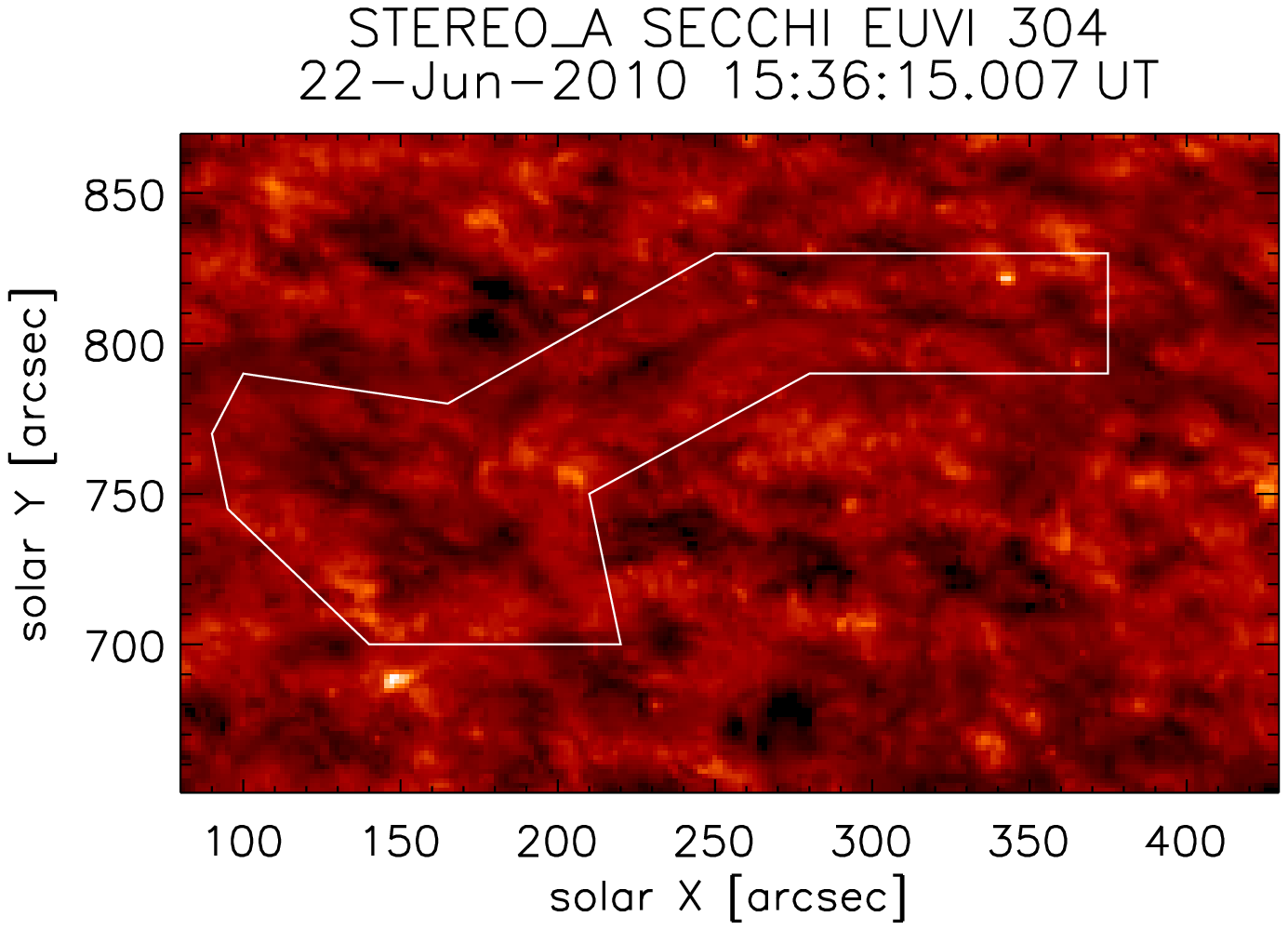}}
\caption{The prominence observed as filament at 304\,\AA\ by the EUVI instrument on\discretionary{-}{-}{-}board of 
the STEREO A satellite. The image was made close in time to XRT and AIA observations. Filament structures are faint and 
geometrically thin so much that area in which the filament 
occurs should be marked by white border. Nevertheless long spine on one end of the filament and two barbs on the other 
are well distinguishable. 
\label{fig:stereo}}
\end{center}
\end{figure}

A quiescent prominence at the North\discretionary{-}{-}{-}West solar limb 
(position around $330\,\deg$) was observed on 22 Jun 2010 by the 
\textit{Solar Optical Telescope} (SOT, \citeauthor{cit:sot} \citeyear{cit:sot}), 
and in soft X rays (SXR) by \textit{X-Ray Telescope} (XRT, \citeauthor{cit:xrt}\ \citeyear{cit:xrt}) 
both on board the Hinode satellite and by the 
SDO/AIA EUV imager. Observations of the prominence 
in the 304, 171 and 193\,\AA\ AIA channels are shown in Fig.~\ref{fig:aiaobs1}. 
Blue rectangles in the images mark the area which was used in calculations 
of the optical thickness $\tau_{193}$ of hydrogen and helium plasma 
by \citet{cit:gunar2014}  which we use in this study. 
This optical thickness should be better denoted as $\tau_{\mathrm{H+He}}(193\,\mathrm{\AA})$ but 
we will use a shorter and more simple name $\tau_{193}$ further in this paper.  
The whole extended prominence is seen in the AIA 304\,\AA\ image,
while only a narrow  vertical  
dark feature is visible in the 193\,\AA\ channel. 
This thin dark structure is seen as well in the AIA 171\,\AA\, but also
extended parts of the prominence are visible in emission in this channel which is the
manifestation of a PCTR (see \citeauthor{parentietal2012}\ \citeyear{parentietal2012}). 
The thin dark structure visible in EUV images from AIA can be identified with the filament spine
seen edge\discretionary{-}{-}{-}on on the limb using observations in 304\,\AA\ channel made by  
the EUVI imager onboard the STEREO A satellite shown in Fig.~\ref{fig:stereo}. 
STEREO A was positioned at such angle (approximately 75\,$\deg$ from Hinode when viewed from Sun) that the prominence 
was seen as a filament. The EUVI observations were made close in time to the AIA observations. On the other hand, 
filament barbs seen in Fig.~\ref{fig:stereo} are most probably extended parts of the prominence. 

\begin{figure*}
\begin{center}
\parbox{0.48\hsize}{
\resizebox{\hsize}{!}{\includegraphics{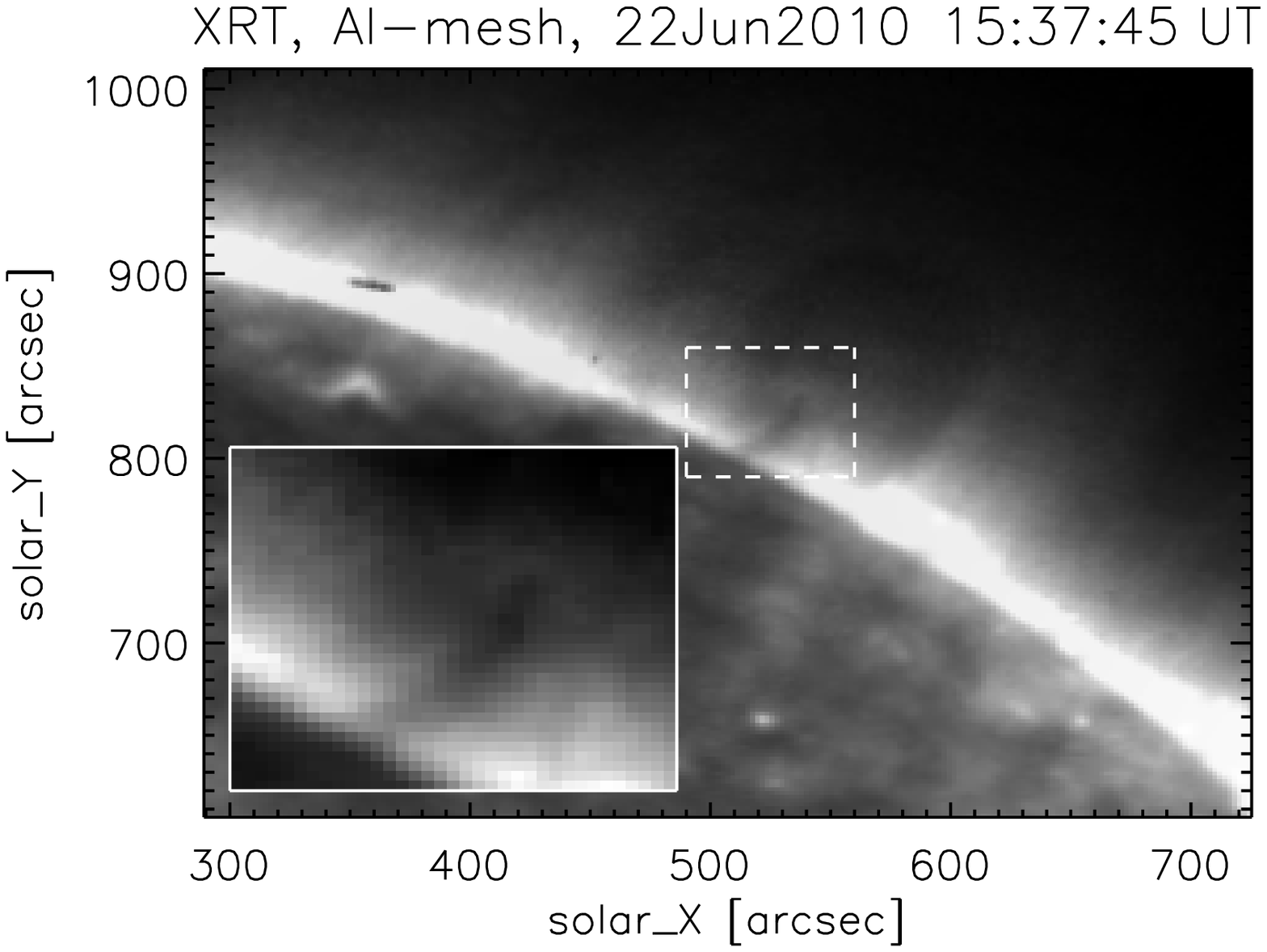}}}\
\parbox{0.48\hsize}{
\resizebox{\hsize}{!}{\includegraphics{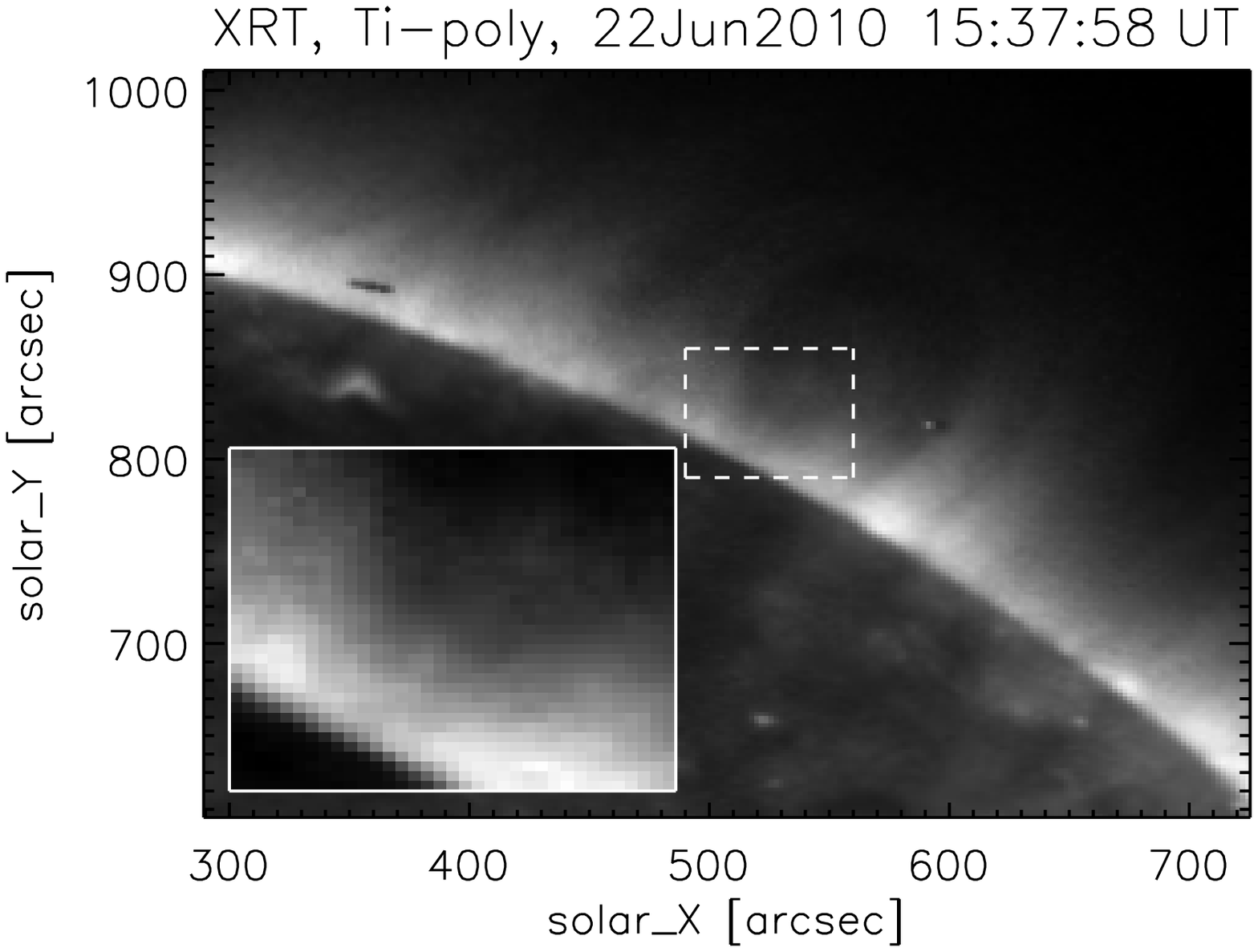}}}
\caption{XRT images 
of the X\discretionary{-}{-}{-}ray corona at place of the prominence and its vicinity made with
the Al\discretionary{-}{-}{-}mesh (the left panel) and Ti\discretionary{-}{-}{-}poly 
(the right panel) focal\discretionary{-}{-}{-}length filters made at 15:37\,UT. Other 
Al\discretionary{-}{-}{-}mesh and Ti\discretionary{-}{-}{-}poly images of the prominence place made 
between 13:18:13 and 17:39:31\,UT look similar as these two images.\label{fig:xrtobs1}}
\end{center}
\end{figure*}

XRT observed the corona at the prominence location and its vicinity in SXR
using two of its focal\discretionary{-}{-}{-}plane analysis filters Al\discretionary{-}{-}{-}mesh 
and Ti\discretionary{-}{-}{-}poly. XRT observations were 
made between 13:18:13 and 17:39:31\,UT with exposure times from 4.1 up to 16.4\,s. Field of view (FOV) of the 
observed images is $788$\,arcsec\,$\times\,788$\,arcsec and dimensions of one pixel is 
$2.06\,$\,arcsec\,$\times\,2.06$\,arcsec. The data were processed using standard data reduction routines in
SolarSoft \citep{cit:solarsoft} provided by the XRT team \citep{cit:xrtdatareduction}. Observations in the two filters made 
at 15:37:45 and 15:37:58\,UT, respectively, are shown in Fig.~\ref{fig:xrtobs1}. 
There is a time\discretionary{-}{-}{-}dependent contamination layer on the CCD (see \citeauthor{cit:xrtspots1}\ 
\citeyear{cit:xrtspots1}, \citeyear{cit:xrtspots2}) and contamination spots which manifest as small dark areas 
in XRT observation images. Especially, the original Al\discretionary{-}{-}{-}mesh image was studded with many such spots. 
In both images in Fig.~\ref{fig:xrtobs1}, spots were retouched by an interpolation to see better 
the darkening occurring at the prominence spine.    
In the Al\discretionary{-}{-}{-}mesh image in the left panel of the figure, dark radial 
structure (seen in the AIA 193\,\AA\ and 171\,\AA\ images) at spine is clearly visible, while
in Ti\discretionary{-}{-}{-}poly it is much weaker. Because the response of both filters to SXR
is rather similar (peaked around 10\,\AA), an additional darkening in the Al\discretionary{-}{-}{-}mesh image
is most probably caused by a secondary peak in its transmittance function - we address this
effect in the present paper. 

\section{Soft X-ray Visibility of Prominences}
\label{s:xrayvisibility}

\begin{figure}
\begin{center}
\resizebox{\hsize}{!}{\includegraphics{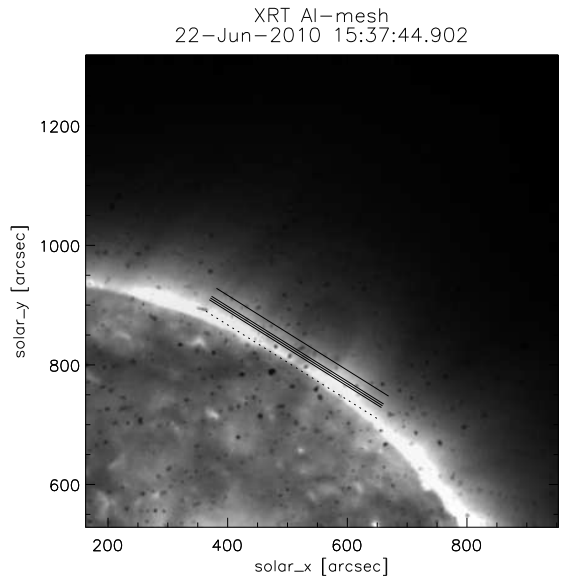}}
\caption{The original (before retouching the contamination spots) XRT Al\discretionary{-}{-}{-}mesh image taken 
at 15:37\,UT. Intensities along cuts made tangentially to limb in four heights are used for estimations 
of depth of depression at the place of prominence spine. The cuts are marked by black solid lines and tangent to 
the limb at the prominence spine by dashed black line. Cuts at the same heights were made also in 
the Ti\discretionary{-}{-}{-}poly image taken almost 13 seconds later.\label{fig:xrtimg4cuts}}
\end{center}
\end{figure}

\begin{figure*}
\begin{center}
\parbox{0.48\hsize}{
\resizebox{\hsize}{!}{\includegraphics{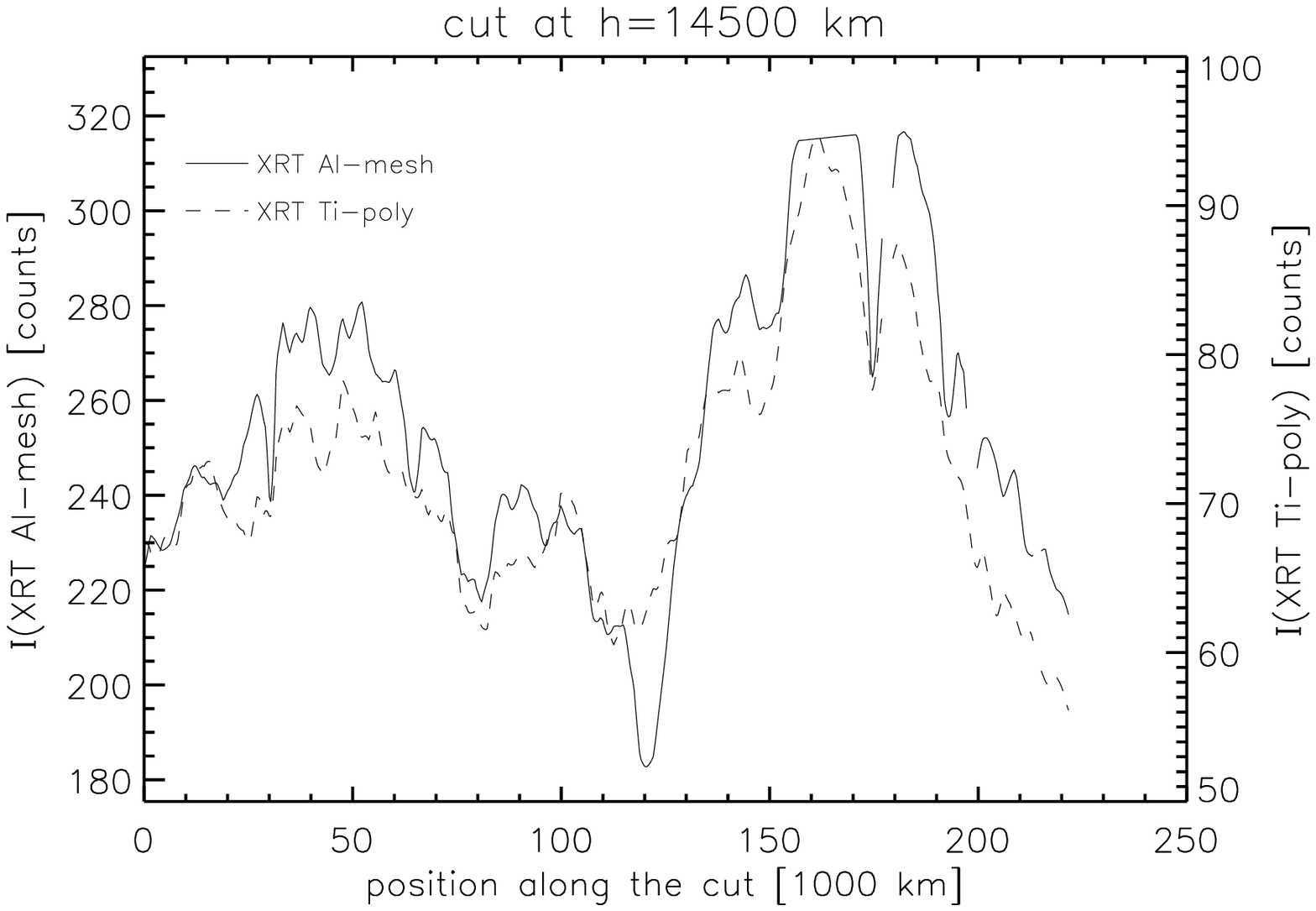}}}\
\parbox{0.48\hsize}{
\resizebox{\hsize}{!}{\includegraphics{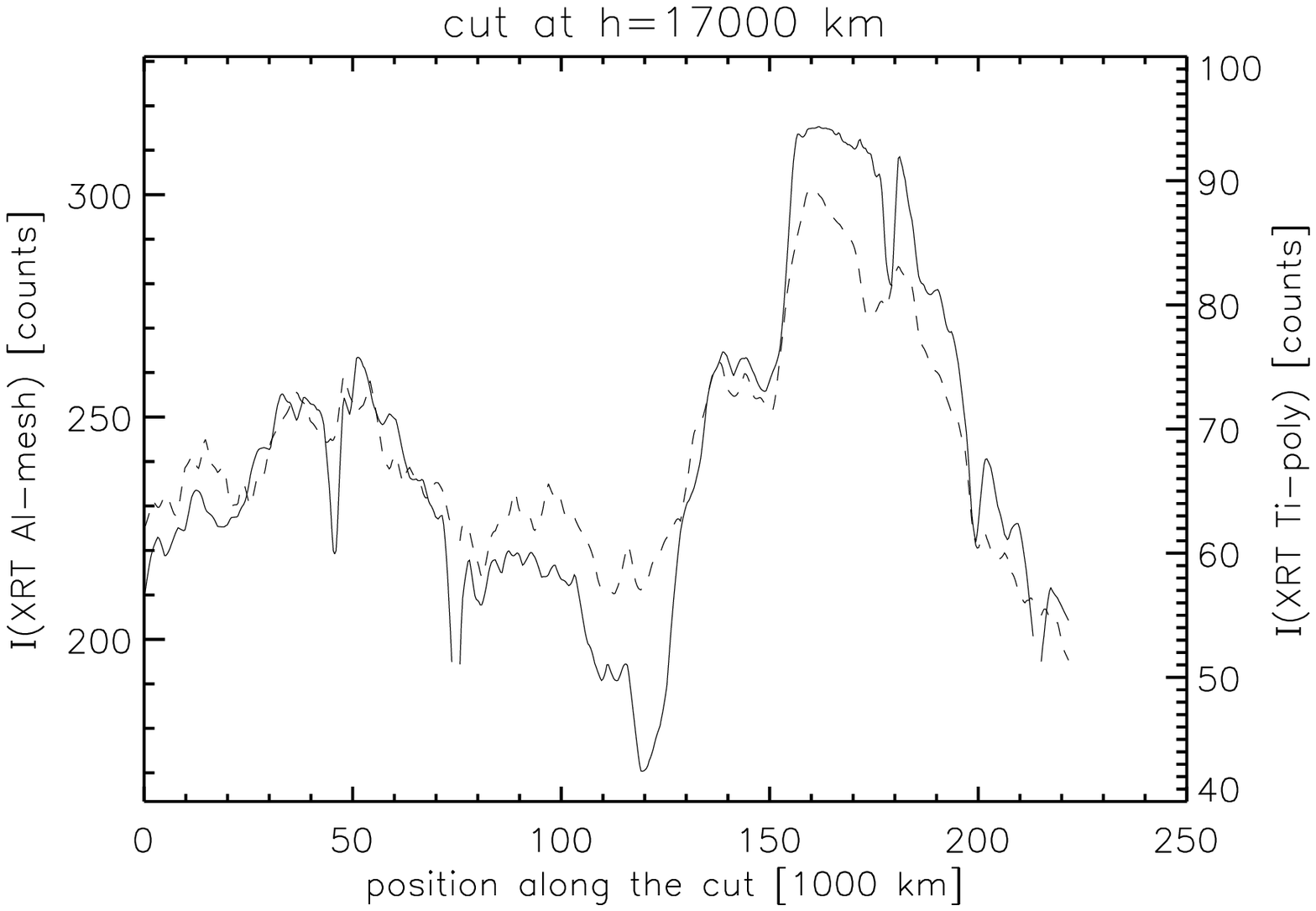}}}\\
\parbox{0.48\hsize}{
\resizebox{\hsize}{!}{\includegraphics{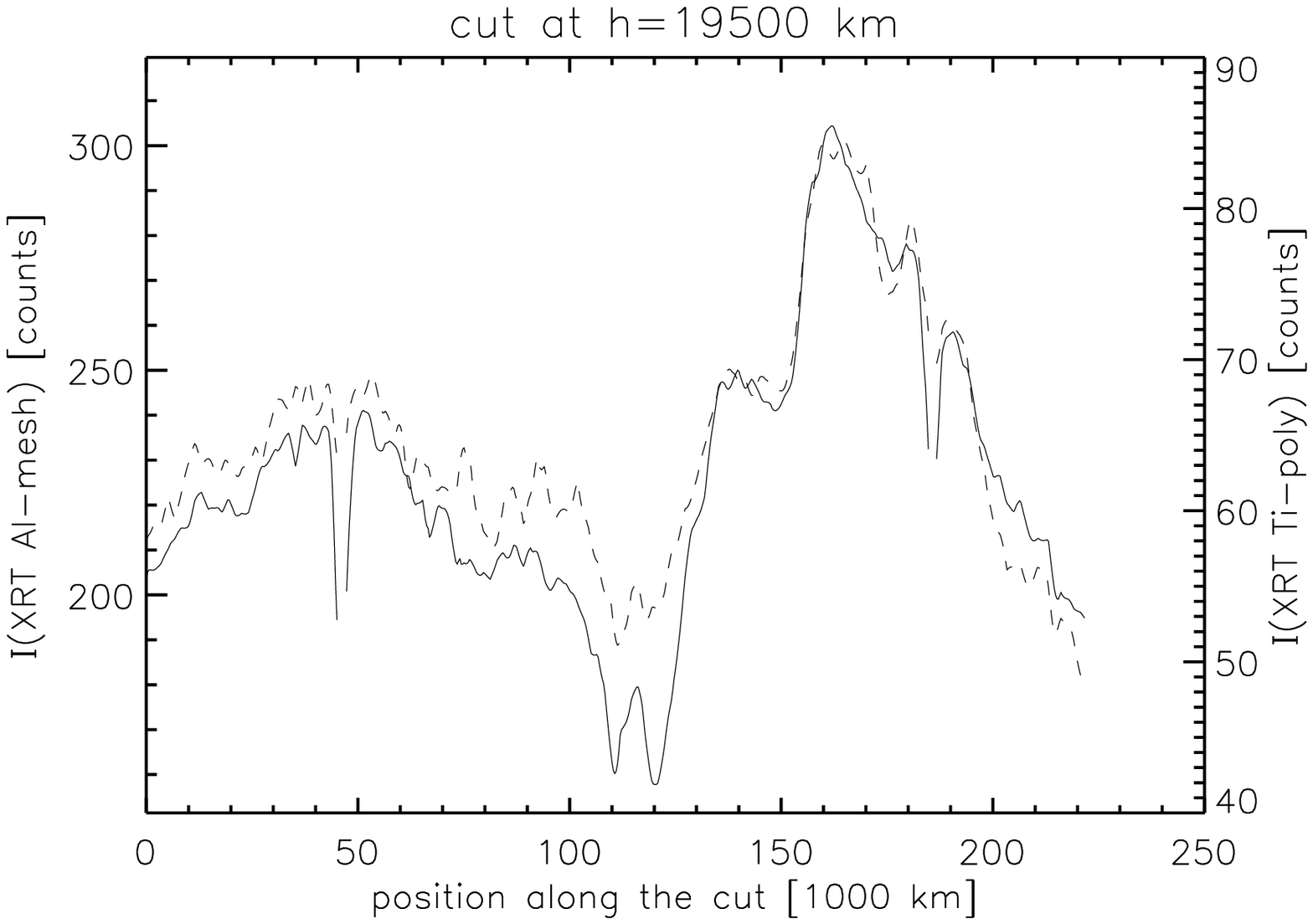}}}\
\parbox{0.48\hsize}{
\resizebox{\hsize}{!}{\includegraphics{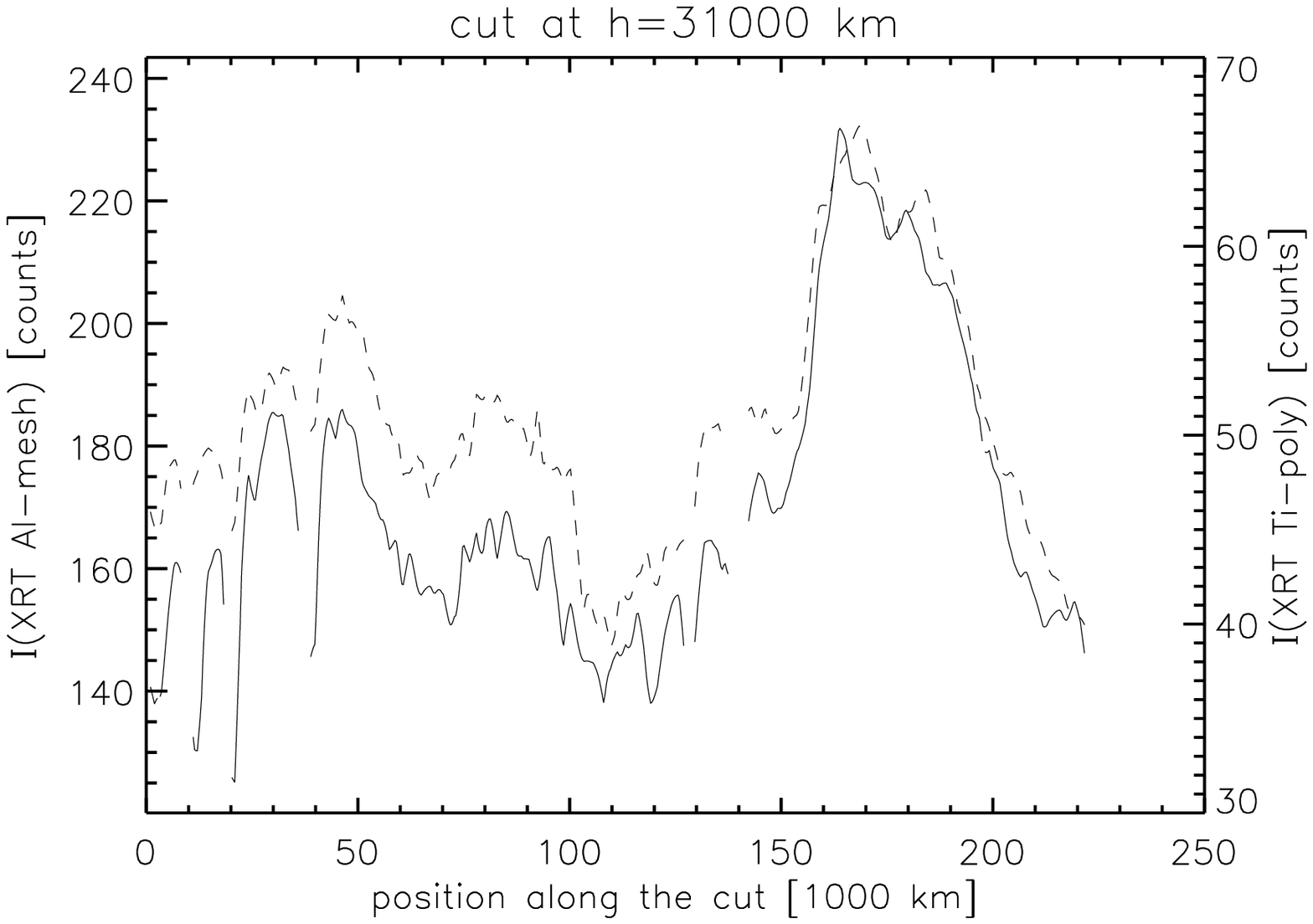}}}
\caption{Intensity distributions along the four cuts made at heights 14\,500, 17\,000, 19\,500 and 31\,000\,km
above the limb in the Al\discretionary{-}{-}{-}mesh and Ti\discretionary{-}{-}{-}poly images taken at 15:37\,UT. 
Depression at the prominence spine (positions along the cuts 110\,000\,--\,120\,000\,km) is seen well in data 
from both filters in all four heights although it is much shallower in the Ti\discretionary{-}{-}{-}poly data. 
For estimating of the quiet\discretionary{-}{-}{-}corona intensity along each cut, an average value at 
positions around 50\,000\,km along the cut outside contamination spots was used. \label{fig:xrtplots4cuts}}
\end{center}
\end{figure*}

In order to investigate darkening in the SXR images
in detail, we made cuts tangentially to the limb in both 
Al\discretionary{-}{-}{-}mesh and Ti\discretionary{-}{-}{-}poly images taken 
at 15:37:45 and 15:37:58\,UT, respectively, at four  
different heights as shown in Fig.~\ref{fig:xrtimg4cuts}. 
Heights above the limb at which the cuts were made were 
chosen so that they would not intersect any contamination spot at least at  
places of the prominence location and its vicinity. Three cuts were made close to 
each other at heights 14\,500, 17\,000 and 19\,500\,km and the fourth one
at a larger height of 31\,000\,km. Resulting intensity plots along cuts are shown 
in Fig.~\ref{fig:xrtplots4cuts}. A noticeable decrease in the
intensity occurs at position of the prominence spine in 
Al\discretionary{-}{-}{-}mesh filter in all four cuts. In Ti\discretionary{-}{-}{-}poly  
along the four cuts a decrease also occurs at position of the dark prominence structure, 
but somewhat shallower than in the case of Al\discretionary{-}{-}{-}mesh data. 

There are two known mechanisms that can be responsible for the darkening: absorption 
of background coronal radiation by the cool prominence plasma and/or the so\discretionary{-}{-}{-}called 
coronal emissivity deficit (formerly called volume or emissivity blocking).
Although \citet{cit:anzeretal2007} already showed that there is a negligible amount of 
absorption in the hydrogen and helium prominence plasma at 
wavelengths around 50\,\AA\, (they used the Yohkoh observations), we can expect some additional opacity due to metals
around 10 \AA\, where both XRT filters have their peaks in the X-ray domain. Moreover, 
the line of sight crosses an extended volume occupied by a cool prominence plasma not emitting in SXR that can 
cause lower intensities in the corona -- the coronal emissivity deficit. 

However, the two filters we used have quite different responses to the EUV part of the spectrum.
While the Ti\discretionary{-}{-}{-}poly transmittance has mainly one peak around 10 \AA,  
the Al\discretionary{-}{-}{-}mesh filter has two transmittance maxima, one also around 10 \AA\, and the 
other one around 171 \AA. This then means that apart from the absorption
and emissivity deficit, a more pronounced darkening in Al\discretionary{-}{-}{-}mesh can be explained by
a contamination from the secondary EUV peak of the filter. In the following subsections we first estimate the
total prominence opacity in the X-ray domain taking into account several important metals and then explain
the darkening in Ti\discretionary{-}{-}{-}poly alone. In the last subsection we show how 
the Al\discretionary{-}{-}{-}mesh images are affected by the secondary EUV transmittance peak.

\section{Mechanisms of prominence SXR darkening}
\label{s:mechsxrdark}

\subsection{Soft X-ray absorption}
\label{ss:xrayabs}

The absorption of X-ray background coronal radiation is caused by hydrogen and helium resonance continua and by continua of some 
metals (the process called photoionisation). The cross sections and total optical thickness at resonance continua of hydrogen and helium 
mixture have been calculated by \citet{cit:ah2005} for cool gas located at the corona. 
Cross sections of neutral hydrogen and singly ionised helium at a given wavelength ($\lambda$) of the resonance continuum are 
proportional to $\lambda^{3}$
\begin{equation}
\sigma_{\mathrm{H I}}(\lambda)= \sigma_0\,g_{\mathrm{H I}}(\lambda)\,\left(\frac{\lambda}{912}\right)^{3} 
\label{eq:csh}
\end{equation}
and
\begin{equation}
\sigma_{\mathrm{He II}}(\lambda)= 16\,\sigma_0~g_{\mathrm{He II}}(\lambda)\,\left(\frac{\lambda}{912}\right)^{3} \, , 
\label{eq:cshe2}
\end{equation}
where 912 presents the Lyman limit of the neutral hydrogen in units of \AA.  
Here $\sigma_{0} = 7.91\times10^{-18}$~cm$^{-2}$, $g_{\rm H I}$ is the hydrogen Gaunt factor (see \citealp{cit:kar61}), 
and $g_{\mathrm{He II}}(\lambda)=g_{\mathrm{H I}}(4\,\lambda)$. The cross section of neutral helium is obtained from Fig.~2 
in \citet{cit:bro70}.

\begin{figure}
\begin{center}
\resizebox{\hsize}{!}{\includegraphics{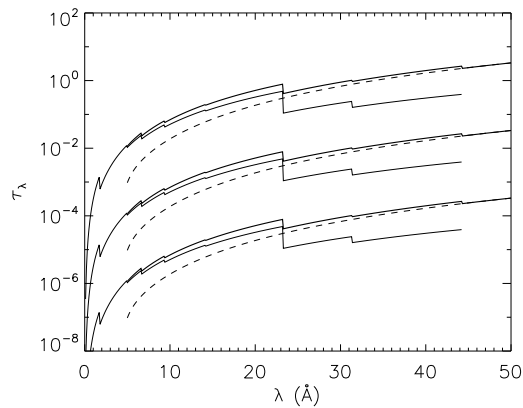}}
\caption{
Plot of the optical thickness due to metals as a function of wavelength marked with
thin solid line. Lower set of curves is for $N_{\rm H}$=$10^{17}\,\mathrm{cm}^{-2}$, middle 
for $N_{\rm H}$=$10^{19}\,\mathrm{cm}^{-2}$ and upper for $N_{\rm H}$=$10^{21}\,\mathrm{cm}^{-2}$.
From 5\,\AA\, we also mark the optical thickness due to partially ionised hydrogen and
helium mixture (with $r_{\rm He}$ = 0.1, $i$ = 0.3, $j_1$ = 0.3, $j_2$ = 0)
marked with thin dashed line. The total contribution of hydrogen, helium and metals between 5
and 50\,\AA\ is marked by thick solid line.}
\label{fig:abs}
\end{center}
\end{figure}

The optical thickness of hydrogen and helium mixture at SXR spectral range (see \citealt{cit:ah2005}) is given by 
\setlength\arraycolsep{1.4pt}
\begin{eqnarray}
\tau_{\mathrm{H+He}}(\lambda) & = & N_{\mathrm{H}}\,\left\{(1-i)\,\sigma_{\mathrm{H I}}(\lambda)+ 
 r_{\mathrm{He}}\,\left[(1-\right.\right. \nonumber \\[-1.5ex]
 &  &  \label{eq:tauhhe} \\[-1.5ex]
 &  &  \left.\left.-j_1-j_2)\,\sigma_{\mathrm{He I}}(\lambda) + 
j_1\,\sigma_{\mathrm{He II}}(\lambda)\right]\right\} \,,\nonumber
\end{eqnarray}
\setlength\arraycolsep{\the\origacs}\noindent
where $N_{\mathrm{H}}$ is the total hydrogen column density ($N_{\mathrm{H}}= N_{\mathrm{H I}}+N_{\mathrm{p}}$), $N_{\mathrm{H I}}$ 
and $N_{\mathrm{p}}$ are neutral hydrogen and  proton column densities, respectively, $i$ is the ionisation degree of hydrogen defined as the ratio 
between the proton and total hydrogen column density, $r_{\mathrm{He}}$ is  the abundance of the helium relative to 
hydrogen ($N_{\mathrm{He}}/N_{\mathrm{H}}$) and ionisation degrees of neutral and singly ionised helium  
$j_1, j_2$, respectively, are defined as $j_1 = N_{\mathrm{He II}}/N_{\mathrm{He}}$\,, 
$j_2 = N_{\mathrm{He III}}/N_{\mathrm{He}}$\,, where $N_{\mathrm{He}} = N_{\mathrm{He I}} + N_{\mathrm{He II}} +N_{\mathrm{He III}}$. 
For three typical values of $N_{\mathrm{H}}$ ($10^{17}, 10^{19}, 10^{21}\,\mathrm{cm}^{-2}$) taken from \citet{cit:gut93},
hydrogen and helium mixture with $r_{\mathrm{He}}$=0.1, $i$=0.3, $j_1$=0.3 and $j_2$=0, the optical thickness 
is calculated between 5 and 50\,\AA\ as 
shown in Fig.~\ref{fig:abs} (thin dashed lines for various $N_{\rm H}$). 

X-ray absorption by hydrogen and helium at 50\,\AA\ was computed already by \citet{cit:anzeretal2007}. Here we extend the 
wavelength range below 50\,\AA\ and add contributions of eight metal continua, {\it i.e.} C, N, O, Ne, Mg, Si, S, Fe.
Photoionisation cross sections of these eight abundant metals are computed using an approximate formula which depends on the  
energy $E$ (see \citealp{cit:lon81}) 
\begin{equation}
\sigma(E) = \sigma_{\mathrm{T}} \left[a\,\left(\frac{E_{\mathrm{T}}}{E}\right)^3 + (1-a)\,\left(\frac{E_{\mathrm{T}}}{E}\right)^4\right]\,.
\label{eq:sig}
\end{equation} 
Here $E_{\mathrm{T}}$ and $\sigma_{\mathrm{T}}$ are the threshold energy and threshold cross section, respectively, 
and $a$ is a parameter chosen to match the 
slope near the threshold (see Table~3 in \citealt{cit:lon81}). Comparison of cross sections of metals between the approximate formula 
given by \citet{cit:lon81} and Fig.~2 in \citet{cit:bro70}, which is mostly cited in the literature, gives a difference of the order 
of 10\,\%. The best agreement is for Si (1\,\%) and the worst for 
S (21\,\%). Optical thickness of metals is expressed in the form
\begin{equation}
\tau_{\lambda} = N_{\mathrm{H}}\,\sum\limits_{i=1}^8 \sigma_i(E)\,A_i\,,
\label{eq:ot}
\end{equation}
where $A_i$ represents the abundance of a chosen metal marked by subscript $i$ given in Table~3 of \citet{cit:lon81}.
Optical thickness is calculated for all eight abundant metal elements under the condition $E\geq E_{\mathrm{T}}$, otherwise 
$\sigma(E)$=0. Note that in our calculations we used the photospheric abundances of metals; the coronal abundances would 
lead to even smaller opacities. Note here that the metallic opacity is not sensitive to the ionisation degree of individual metals,
because we are dealing here only with the inner-shell electrons (see also \citealt{cit:lon81}).

Fig.~\ref{fig:abs} presents the optical thickness versus wavelength below 50\,\AA\ for three values of $N_{\mathrm{H}}$. 
Contribution of metals is marked with thin solid line. Small "jumps" on the curves occur when metals 
stop to contribute above certain wavelength (their $E_{\mathrm{T}}$). The total contribution of hydrogen and helium 
mixture and of all metals is presented by thick solid lines. Above 20\,\AA\ 
the contribution of metals is negligible compared to hydrogen and helium mixture. It is well seen that 
total $\tau_{\lambda}$ (hydrogen, helium and metals all together) is practically negligible in the SXR domain.

From the amount of EUV coronal emission at wavelengths below 912\,\AA\
absorbed by the hydrogen and helium prominence plasma, \citet{cit:kucera98} and \citet{cit:golub99} 
found the hydrogen column density $10^{18}$\,--\,$10^{19}$\,$\mathrm{cm}^{-2}$
in quiescent prominences. The hydrogen column densities in the same range were estimated also by 
\citet{schwartz2014} from observations of six quiescent prominences in EUV, SXR and H$\alpha$.
Even if one considers a limiting value 10$^{21}$ \,$\mathrm{cm}^{-2}$
which, for example, would correspond to hydrogen density 10$^{11}$\,$\mathrm{cm}^{-3}$ and the prominence
extension of 10$^5$ km,
the optical thickness around 10\,\AA\ (where both Al\discretionary{-}{-}{-}mesh and 
Ti\discretionary{-}{-}{-}poly filters have the maximum responsibility) is below 0.02.
Therefore the absorption mechanism can not explain the observed darkening
and we must turn our attention
to the effect of emissivity deficit.

\subsection{Emissivity Deficit}
\label{ss:emisdef}

In EUV and SXR, the prominence will appear dark in the coronal line/continuum emitted at
temperatures higher than 10$^6$\,K. We assume that this is due to the absorption and emissivity deficit,
i.e.
\begin{equation}
I_{\mathrm{prom}}(\lambda)=I_{\mathrm{fg}}(\lambda)+I_{\mathrm{bg}}(\lambda)\,\exp\left(-\tau_{\lambda}\right)\,,
\label{eq:i_prom_ced1}
\end{equation} 
where $I_{\mathrm{fg}}(\lambda)$ and $I_{\mathrm{bg}}(\lambda)$ are intensities of the radiation emitted by the corona in front and beyond 
the prominence, respectively. Assuming the most simple case when these intensities are equal (symmetric corona),
we can write

\begin{equation}
I_{\mathrm{prom}}(\lambda)=I_{\mathrm{c}}(\lambda)\,\left[1+\exp\left(-\tau_{\lambda}\right)\right]\,,
\label{eq:i_prom_ced2}
\end{equation} 
where
$I_{\mathrm{c}}\equiv I_{\mathrm{fg}}=I_{\mathrm{bg}}$.

In this paper we express the prominence darkening in terms of the {\em intensity ratio R} which we define as

\begin{equation}
R = \frac{I_{\mathrm{prom}}}{I_0} \, ,
\end{equation}
where $I_0$ is the coronal intensity measured close to the prominence. In the case of a negligible absorption,
$R=2 I_{\rm c}/I_0$ which demonstrates the effect of emissivity deficit, i.e. the lack of hot coronal emission at the 
volume occupied by the cool
prominence material - see below. 
Sometimes it is also useful to express the prominence darkening in terms of the {\em contrast}, which can be defined
as $C=1 - R$.
$C$ is zero in the case of no prominence visibility. If there is a negligible absorption,  $C=1-2I_{\rm c}/I_0$ and, on the
other hand, 
for large $\tau$, $C=1-I_{\rm c}/I_0$. In case of $I_{\rm c}=I_0/2$ (no deficit effect) the latter
will give $C=1/2$.

The coronal intensity $I_{\rm c}$ at the prominence location 
is obtained by integration of the coronal emissivity along the line of sight (LOS), from middle of
the (symmetrical) cool structure positioned at the limb to coronal boundaries 

\begin{equation}
I_{\mathrm{c}}(\lambda)=\int_{0}^{\infty}\,C_{\lambda}(n_e,T)\,\frac{n(\mathrm{H})}{n_{\rm e}}\,n_{\rm e}^2\,
dl\, ,
\label{eq:ed2}
\end{equation} 
where $l$ is the position along the LOS expressed as 

\begin{equation}
l=\sqrt{r^2-\left(R_{\mathrm{Sun}}+h\right)^2}\,.
\label{eq:posalonglos}
\end{equation} 
Here $r$ is radial position in the corona,
$R_{\mathrm{Sun}}$ the solar radius and $h$ the height above the limb. 
$n_{\rm e}$ and $n_{\rm H}$ are the electron and hydrogen densities, respectively, $T$ is the kinetic temperature
and $C_{\lambda}$ is the contribution function calculated using the statistical equilibrium and CHIANTI atomic database 
version 7 \citep{cit:chianti1, cit:chianti2}. For the ratio $n(\mathrm{H})/n_{\rm e}$
a common coronal value 0.83 is adopted. Distributions of the temperature and electron density
with height above the solar surface in the quiet corona were taken from \citet{cit:qstemp} and 
\citet{cit:qseldens}, respectively. In Eq.~(\ref{eq:ed2}) we first integrate from the middle
of the cool prominence structure up to its boundary $D_{\mathrm{geom}}/2$ 
(where $D_{\mathrm{geom}}$ represents the total LOS extension of the 
prominence), this  
automatically accounts for the emissivity deficit because $C_{\lambda}$ is there essentially zero. From the coronal
part we get actual $I_{\rm c}$. In the quiet corona outside the prominence, this integral gives simply $I_0/2$.
Finally, the signal measured by XRT is calculated by the integration of $I(\lambda)$ multiplied by the filter 
response function $f(\lambda)$ over the wavelength range of the filter.

\begin{equation}
E=\int \,I(\lambda)\,f(\lambda)\,d\lambda\,,
\label{eq:xrtsignal}
\end{equation}
where $I(\lambda)$ is either $I_{\rm prom}$, $I_{\rm c}$ or $I_0$. Response functions $f(\lambda)$ for both 
Al\discretionary{-}{-}{-}mesh and Ti\discretionary{-}{-}{-}poly filters are shown in plots in 
Fig.~\ref{fig:iqs_almesh_tipoly}.  

Length 1.8$\times10^5$\,km of the prominence spine in 
projection on the solar disc was measured in image Fig.~\ref{fig:stereo} made at 304\,\AA\ with the EUVI instrument 
onboard the STEREO A which observed the prominence as filament. But real length of the spine can be larger. Moreover, 
it can be possible for the prominence on\discretionary{-}{-}{-}limb observations that the line of sight was not passing 
along whole length of the spine or part of the spine could be hiden behind the limb. Therefore, length estimated according 
to STEREO A observations cannot be used as $D_{\mathrm{geom}}$. For correct derivation of $D_{\mathrm{geom}}$ view from minimum 
three angles is necessary and only two are available (edge\discretionary{-}{-}{-}on viewed from Earth direction and from 
above). Unfortunately, STEREO B was positioned by 75\,$\deg$ from Earth in opposite direction than STEREO A and therefore 
the whole prominence was behind the limb for STEREO B. Thus, we used Ti\discretionary{-}{-}{-}poly observations 
to estimate $D_{\mathrm{geom}}$ at four heights where the cuts have been made. 
$D_{\mathrm{geom}}$ was optimized in order to achieve the best fit
between the observed and computed ratio $R$. Note that in the case of Ti\discretionary{-}{-}{-}poly filter,
the latter is equal to $2E_{\rm c}/E_0$ because the absorption at 10\,\AA\ is considered to be quite negligible and 
contribution of EUV radiation to measured signal is under 1\,\%. The resulting $D_{\mathrm{geom}}$ are shown in 
Table~\ref{tab:compcontr_withdgeom} and is used in the next subsection to evaluate
the darkening in the Al\discretionary{-}{-}{-}mesh images where the filter has a secondary peak in EUV which
contributes to the integral in Eq.~(\ref{eq:xrtsignal}). All resulting values of $D_{\mathrm{geom}}$ are 
of order of magnitude of $10^5$\,km that is close to length of the spine measured in EUVI image 
(Fig.~\ref{fig:stereo}) from the STEREO A satellite. But it has to be also noted that any of them do not 
exceed the spine length as measured in the STEREO A image. 
\begin{deluxetable}{ccccccc}
\tablewidth{0pt}
\tablecaption{Comparison of observed and calculated intensity ratios ($R$) at the prominence spine for 
the Al\discretionary{-}{-}{-}mesh and Ti\discretionary{-}{-}{-}poly filters.
\label{tab:compcontr_withdgeom}}
\tablehead{
\colhead{$h$} & \multicolumn{2}{c}{$R$ from observations} & \colhead{$D_{\mathrm{geom}}$} & \colhead{maximum $\tau_{193}$} & 
\multicolumn{2}{c}{$R$ from calculations} \\
\colhead{$\left[\mathrm{km}\right]$} & \colhead{Ti-poly} & \colhead{Al-mesh} & \colhead{$\left[10^4\,\mathrm{km}\right]$} & 
  & \colhead{Ti-poly} & \colhead{Al-mesh}
 }
\startdata
14\,500 & 0.83 & 0.77 &  $\ 7.8$ & 2.01 & 0.83 & 0.80 \\
17\,000 & 0.82 & 0.79 &  $\ 8.3$ & 2.67 & 0.82 & 0.79 \\ 
19\,500 & 0.81 & 0.76 &  $\ 8.9$ & 2.13 & 0.81 & 0.78 \\
31\,000 & 0.78 & 0.78 &  $10.0$  & 1.40 & 0.78 & 0.76 \\ 
\enddata
\end{deluxetable}

\subsection{EUV Contribution from the Al-mesh Secondary EUV Peak}
\label{ss:almeshsidepeak} 

\begin{figure*}
\parbox{0.48\hsize}{
\resizebox{\hsize}{!}{\includegraphics{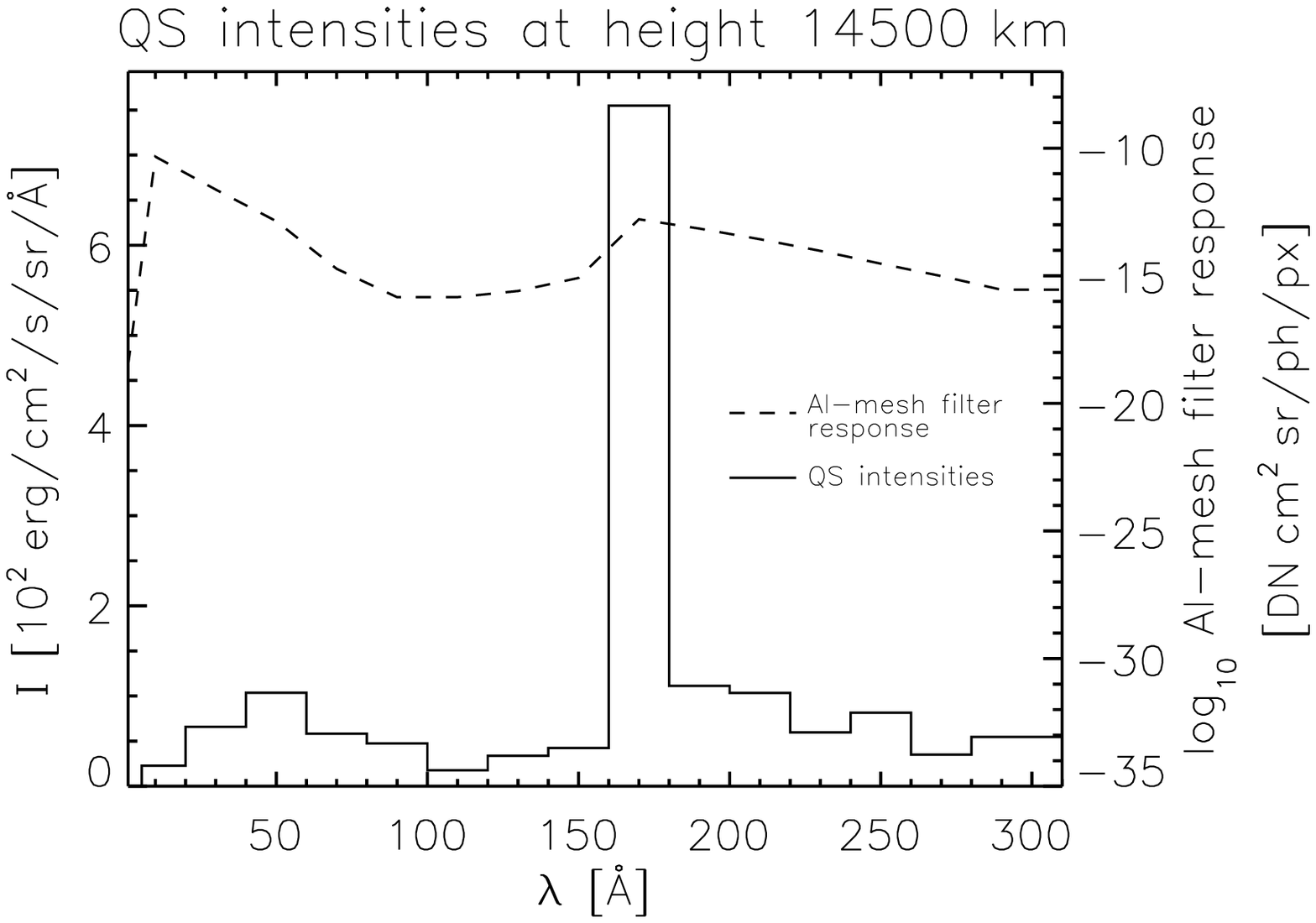}}}\
\parbox{0.48\hsize}{
\resizebox{\hsize}{!}{\includegraphics{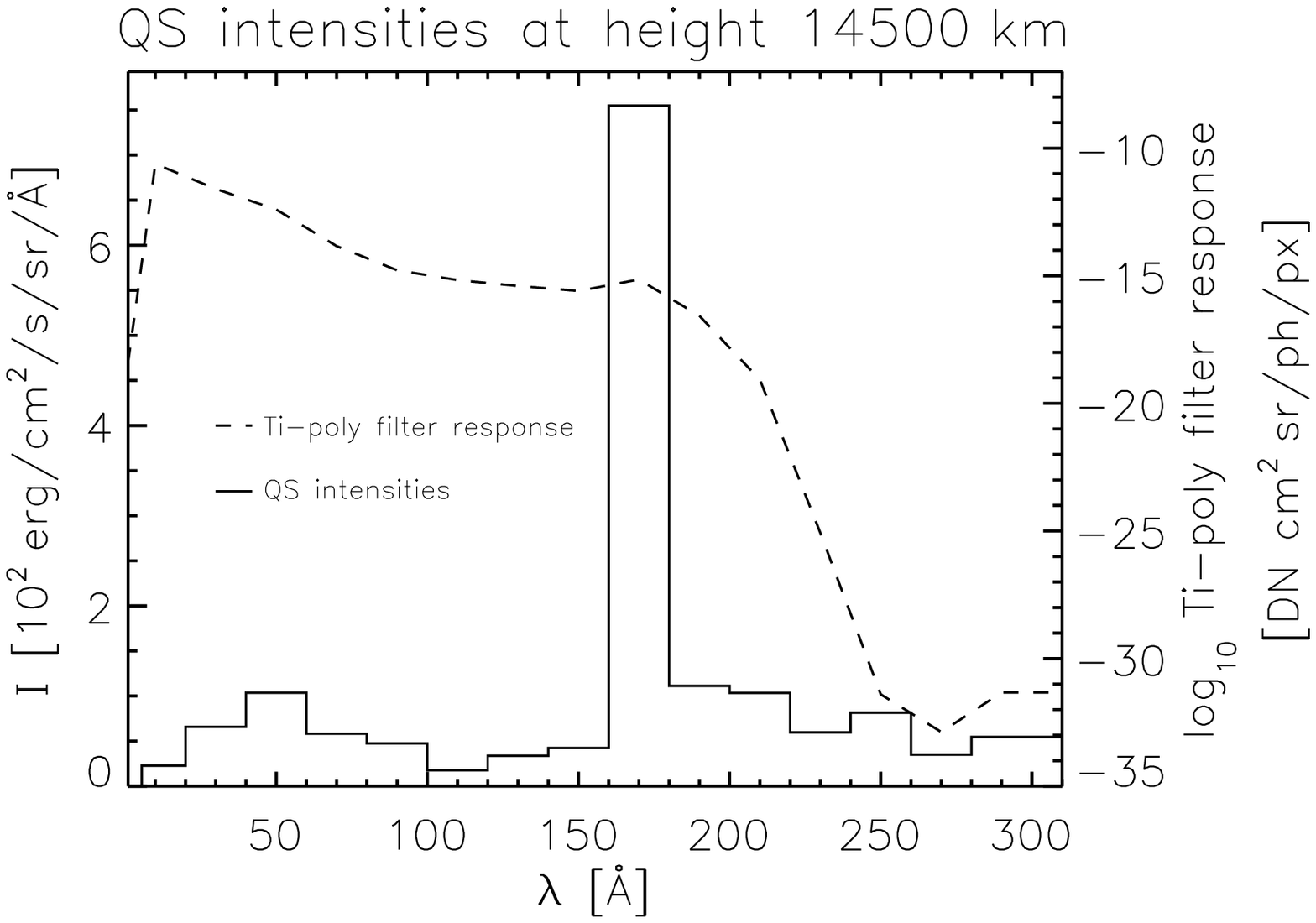}}}
\caption{Intensities of the quiet corona within interval of wavelengths 1\,--\,300\,\AA\ calculated 
using Eq.~(\ref{eq:ed2}) for height 14\,500\,km above the limb. Plotted intensities are mean values 
of spectral intensities within 20\,\AA\ interval bins. Al\discretionary{-}{-}{-}mesh (left panel) and 
Ti\discretionary{-}{-}{-}poly (right panel) filter responses are also plotted. Although the response 
in EUV peak at 170\,\AA\ is almost thousand times lower than for X rays at 10\,\AA\ for Al\discretionary{-}{-}{-}mesh, 
intensities around 170\,\AA\ are approximately $40$ times as large as those at 10\,\AA. Thus, contribution 
of the EUV radiation from wavelengths around 170\,\AA\ to the measured signal is around 11\,\% that cannot 
be neglected. Similar EUV contribution was calculated for the quiet corona also at heights 17\,000 and 19\,500\,km 
while at height 31\,000\,km the contribution was only 9\,\%. In the other hand, the response at 170\,\AA\ for 
the Ti\discretionary{-}{-}{-}poly filter is four orders of magnitude lower than at 10\,\AA, thus EUV contribution 
to the measured signal is under 1\,\%. \label{fig:iqs_almesh_tipoly}}
\end{figure*}

\begin{figure}
\begin{center}
\resizebox{\hsize}{!}{\includegraphics{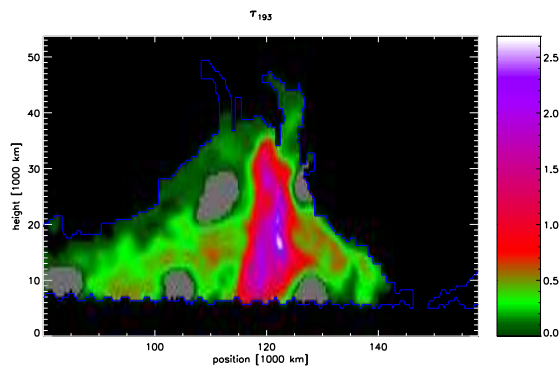}}
\caption{Map of the $\tau_{193}$ optical thickness obtained by  \citet{cit:gunar2014} from 
amount of radiation from behind the same prominence in resonance continua of hydrogen and helium 
assuming symmetrical distribution of coronal emissivity along the line of sight. 
For calculations the method of \citet{schwartz2014} and AIA 193\,\AA\ and XRT Al\discretionary{-}{-}{-}mesh data shown in 
Figs~\ref{fig:aiaobs1} and \ref{fig:xrtimg4cuts}, respectively, were used. 
Gray areas mark positions of contamination spots in the Al\discretionary{-}{-}{-}mesh image 
where the data are corrupted. \label{fig:tau193}}
\end{center}
\end{figure}

\begin{figure}
\begin{center}
\resizebox{\hsize}{!}{\includegraphics{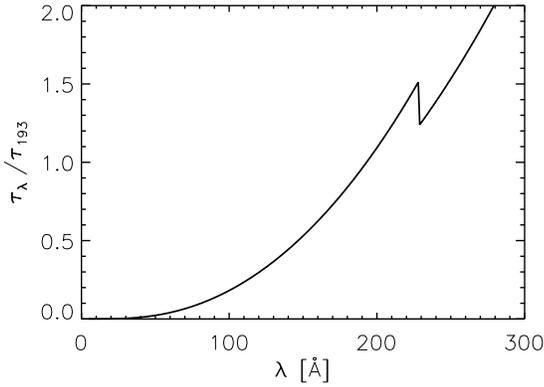}}
\caption{The $\tau_{\mathrm{H+He}}(\lambda)/\tau_{193}$ ratio plotted within wavelength 
range 1\,--\,300\,\AA\ (from hard X rays up to EUV). Values of the optical thickness 
$\tau_{\mathrm{H+He}}(\lambda)$ were computed using the theoretical formula Eq,~(\ref{eq:tauhhe}) 
for $\tau_{\mathrm{H+He}}(\lambda)$ calculation for the common solar helium abundance of $0.1$ 
and ionisation degrees of helium and hydrogen $i$=0.3, $j_1$=0.3 and $j_2$=0. \label{fig:tratio193}}
\end{center}
\end{figure}

To evaluate the intensity ratio $R$ in the case of Al-mesh filter, we proceed in the same
way as in previous subsection. The only difference is that we can not neglect the absorption
because in the EUV domain around 170\,\AA, where the secondary peak of the Al\discretionary{-}{-}{-}mesh filter 
contributes to the measured signal (see Fig.~\ref{fig:iqs_almesh_tipoly}), 
the absorption of coronal radiation by cool hydrogen and helium prominence plasma can be significant. 
Although response for the Al\discretionary{-}{-}{-}mesh filter around 170\,\AA\ is almost three orders of magnitude 
lower than at 10\,\AA, as shown in Fig.~\ref{fig:iqs_almesh_tipoly}, the quiet corona in EUV around 170\,\AA\ is approximately 
$40$ times as intensive as its X\discretionary{-}{-}{-}ray radiation at 10\,\AA. Thus, contribution from EUV to  
signal measured using the Al\discretionary{-}{-}{-}mesh filter at the quiet corona cannot be neglected. And subsequently, 
decrease of this contribution due to the absorption at the prominence can have remarkable impact on the intensity ratio $R$. 
The map of the optical thickness $\tau_{193}$ at 193\,\AA\ of the prominence 
studied here has been already calculated by \citet{cit:gunar2014} using the SDO/AIA observations in the 193\,\AA\ channel 
(upper right panel of Fig.~\ref{fig:aiaobs1}), SXR data from XRT (Fig.~\ref{fig:xrtimg4cuts}) and the method of 
\citet{schwartz2014}. Because of the assumption of symmetrical distribution of the coronal emissivity, factor of the 
coronal asymmetry $\alpha$ equal to 0.5 was used. The $\tau_{193}$ map is shown in Fig.~\ref{fig:tau193}. 
Position and shape of an area with $\tau_{193}$ above 2 corresponds well to the dark radial structure of prominence visible 
in the AIA 193\,\AA\ and XRT Al\discretionary{-}{-}{-}mesh images. Thus, the minimal $R$ of the XRT data for both filters 
corresponds well with the maximum $\tau_{193}$ at all heights above the limb. They can be 
transformed to other wavelengths by multiplying with 
the  $\tau_{\mathrm{H+He}}(193\,\mathrm{\AA})/\tau_{193}$ ratio obtained from Eq.~(\ref{eq:tauhhe}) for estimation 
of the theoretical optical thickness of hydrogen and helium plasma $\tau_{\mathrm{H+He}}(193\,\mathrm{\AA})$, 
see e.g. \citet{cit:ah2005}. In calculations of $\tau_{\mathrm{H+He}}(193\,\mathrm{\AA})$ in wavelength range 
from 1 to 300\,\AA\ (from X rays up to EUV) we adopted the same values of helium abundance and ionisation 
degrees ($r_{\mathrm{He}}$=0.1, $i$=0.3, $j_1$=0.3 and $j_2$=0) as in section~\ref{ss:xrayabs}. 
Resulting ratio is plotted in Fig.~\ref{fig:tratio193}. 
Maximum values of $\tau_{193}$ occurring at the prominence spine are within an interval 1.4\,--\,2.7 for 
heights 14\,500\,--\,31\,000\,km what corresponds to optical thickness at wavelengths around 170\,\AA\ 
of approximately 0.8\,--\,2.2. Such an optical thickness produces remarkable decrease of intensity and subsequently smaller 
EUV contribution (only 7\,--\,8\,\%) to signal measured at the prominence spine using the Al\discretionary{-}{-}{-}mesh 
filter than for the quiet corona (the contribution of 11\,\% at heights 14\,500\,--\,19\,500\,km, 9\,\% at the height 
31\,000\,km). Then again the signal $E_{\mathrm{prom}}$ registered at the prominence by the Al\discretionary{-}{-}{-}mesh
filter is calculated by integration along the wavelength of $I_{\mathrm{prom}}(\lambda)$ multiplied by the 
instrument response $f(\lambda)$ (see Fig.~\ref{fig:iqs_almesh_tipoly}), similarly as in the case of Ti-poly.
Finally, the theoretical intensity ratio $R=E_{\mathrm{prom}}/E_{\mathrm{0}}$ at 
the prominence position is calculated. Note that the emissivity deficit is properly accounted for
by using the values of $D_{\mathrm{geom}}$ obtained in the previous subsection.

For Al\discretionary{-}{-}{-}mesh filter we compare the observed values of $R$ with those calculated assuming the EUV 
contamination in Table~\ref{tab:compcontr_withdgeom}. For values of $D_{\mathrm{geom}}$ of the order of 
$10^5$\,km and maximal $\tau_{193}$ between 1.4 and 2.67 a good agreement between calculated
and observed values of $R$
was achieved for the four selected heights above the limb - see Table 1.. 



\section{Discussion and Conclusions}
\label{s:discussion}
In this paper we studied SXR visibility of the prominence observed 
on 22 June 2010 during the coordinated campaign. We noticed that a dark structure resembling the prominence spine is well visible in 
Hinode/XRT images obtained with Al\discretionary{-}{-}{-}mesh and Ti\discretionary{-}{-}{-}poly 
filters. Positions of the dark structure at all heights above the limb correspond to the 
maximal $\tau_{193}$ as estimated by \citet{cit:gunar2014} for the same
prominence. We examined three possible mechanisms of SXR
prominence darkening: absorption of X rays around 10\,\AA\ by the resonance continua of hydrogen, 
helium and selected metals, influence of the coronal emissivity deficit and the effect of a contamination by
the secondary EUV peak in the case of the Al\discretionary{-}{-}{-}mesh filter. 

Comparison was made for four heights 
above the limb -- three close to each other cutting the prominence spine somewhere in the middle between its bottom (at the limb) 
and top  and the fourth close to its top. For the theoretical calculations, distributions of electron densities and 
temperature in the quiet corona were used and for the absorption of EUV radiation by hydrogen and helium plasma, maximal $\tau_{193}$ values for 
the four heights calculated for this prominence by \citet{cit:gunar2014} were scaled. Then, calculated intensities were integrated 
along wavelength using the  XRT filter responses in order to obtain signal that should be measured by the XRT instrument. 

We found that both absorption in resonance continua of hydrogen and helium of EUV radiation 
that contaminates SXR data and EUV emissivity deficit would lower $R$ at the prominence spine when using the Al\discretionary{-}{-}{-}mesh filter. 
In case of the Ti\discretionary{-}{-}{-}poly filter, lowering of $R$ due to absorption of EUV radiation is  negligible.  

As for absorption of X\discretionary{-}{-}{-}ray radiation by prominence hydrogen and helium  plasma, it is totally negligible 
at 10\,\AA. But when also continua of other elements (metals) such as C, N, O, Ne, Mg, Si, S, and Fe are taken into account, total optical 
thickness cannot be neglected for hydrogen column density exceeding 
$10^{21}\,\mathrm{cm}^{-2}$.  For the prominence studied here we estimated
the hydrogen column density at hydrogen and helium ionisation degrees $i$=0.6, $j_1$=0.5 and $j_2$=0 from the $\tau_{193}$ map 
constructed by \citet{cit:gunar2014} and we found a maximum column density of hydrogen being of 
$2\times10^{19}\,\mathrm{cm}^{-2}$ for this prominence. It means that absorption of X rays at 10\,\AA\ can be neglected for 
the prominence studied here. Only when assuming that the line of sight at the prominence location is passing through a volume 
occupied by the cool prominence plasma not emitting in EUV and X\discretionary{-}{-}{-}rays, the contrast comparable to 
observations is achieved due to the coronal emissivity deficit. For simplicity position of the prominence exactly at the limb 
was assumed. For the geometrical thickness $D_{\mathrm{geom}}$ of the prominence spine along the line of sight of the order 
of $10^5\,\mathrm{km}$, calculated intensity ratios comparable to those obtained from observations were obtained for all four 
selected heights (Table~\ref{tab:compcontr_withdgeom}) for both Al\discretionary{-}{-}{-}mesh and 
Ti\discretionary{-}{-}{-}poly filters. Although $D_{\mathrm{geom}}$ is increasing with height, its variations are not exceeding 
20\,\%. Thus, it can be just due to noise in the XRT Ti-poly data. But increase of $D_{\mathrm{geom}}$ with height could be also 
a real effect of the prominence shape. Unfortunately, observations of the prominence only in two viewing angles are 
available -- edge\discretionary{-}{-}{-}on on the limb and viewed as a filament projected on the disk from STEREO A. 
Therefore it is not possible to infer reliably its 3D shape and subsequently geometrical thickness of the prominence at 
the four heights. Thus, it is not possible to distinguish whether increase of $D_{\mathrm{geom}}$ with height is caused by 
noise in the data or by shape of the prominence. Although, such behaviour of the geometrical thickness might conform a 
loop\discretionary{-}{-}{-}like shape of the prominence where at smaller heights the line of sight is passing through its 
vertical parts while in larger heights the line of sight is passing along its horizontal part. In case of the the 
Al\discretionary{-}{-}{-}mesh filter also the EUV transmittance peak that contaminates the measured signal was taken into 
account. Calculated EUV contribution to the signal in case of the Al\discretionary{-}{-}{-}mesh filter 
for the quiet corona is 11\,\% at heights 14500--19500 km. The EUV contribution at the prominence spine
is 7--8\,\% in all four heights. This difference in EUV contributions causes decrease of 
measured XRT signal at the prominence spine together with the emissivity deficit. Comparing contribution to the signal 
from quiet\discretionary{-}{-}{-}Sun radiation in X rays (main peak of the Al\discretionary{-}{-}{-}mesh filter transmittance 
within wavelength interval 1\,--\,30\,\AA) and EUV (secondary peak at 160\,--\,210\,\AA) it was found
that contribution of EUV to depression of measured signal in the quiet corona decreases steeply with height.
At the height of 31\,000\,km the EUV contribution to the signal in the quiet corona is lower -- only 9\,\% while 
at the prominence spine the contribution is the same as at lower heights. Thus, lower EUV contribution to the measured
signal in the quiet corona at $h$=31\,000\,km causes notably less dramatic intensity decrease at the 
prominence for this height than at lower heights as can be seen in Fig.~\ref{fig:xrtplots4cuts}. 
Although mainly the coronal emissivity deficit is responsible for visibility of this prominence in XRT images, in case 
of the Al\discretionary{-}{-}{-}mesh filter a fraction of 16\,--\,25\,\% of the total darkening comes from the EUV 
contamination. Therefore depression of the measured signal at the prominence in case of the Al\discretionary{-}{-}{-}mesh 
filter is more prominent and its variations with height are larger than in case of the Ti\discretionary{-}{-}{-}poly filter. 
In contrast, the  contamination of Ti\discretionary{-}{-}{-}poly signal by the EUV radiation is negligible and thus the 
emissivity deficit only causes depression at the prominence spine in XRT observations made with the 
Ti\discretionary{-}{-}{-}poly filter.    
\acknowledgments
P.S. and P.H. acknowledge the support from grant P209/12/0906 of the Grant Agency of the Czech Republic. 
Work of P.S. and P.H. was supported by the project RVO:\,67985815. Work of P.S. was also supported by the 
grant project VEGA\,2/0108/12 of the Science Grant Agency. 
P.S. acknowledges support from the Slovak Research and Development Agency under the contract No. APVV-0816-11.
S.J. and P.H. acknowledge the hospitality of the Astronomical Institute of the Slovak Academy of Sciences 
and University of Ljubljana during the course of this investigation. 
We are thankful to Dr.~E.E. DeLuca for useful discussions concerning the XRT observations used in this work.  
The AIA data are courtesy of NASA/SDO and the AIA science team. CHIANTI is a collaborative project involving 
George Mason University, the University of Michigan (USA) and the University of Cambridge (UK).
\end{document}